# Atmospheric Temperature anomalies as manifestation of the dark Universe

**K. Zioutas[1], V. Anastassopoulos[1], A. Argiriou[1], G. Cantatore[2], S. Cetin[3], H. Fischer[4], A. Gardikiotis[1,5], H. Haralambous[6], D. H. H. Hoffmann[7], S. Hofmann[8], M. Karuza[9], A. Kryemadhi[10], M. Maroudas[1#&], A. Mastronikolis[11], C. Oikonomou[6], K. Ozbozduman[12], Y. K. Semertzidis[13,14]**

[1] Physics Department, University of Patras, 26504 Patras-Rio, Greece
[2] Physics Department, University & INFN of Trieste, 34127 Trieste, Italy
[3] Institute of Sciences, Istinye University, 34396, Istanbul, Türkiye
[4] Physikalisches Institut, Albert-Ludwigs-Universität Freiburg, 79104 Freiburg, Germany
[5] INFN, Sezione di Padova, Padova, Italy
[6] Frederick University, Electrical Engineering Department, Nicosia, Cyprus
[7] School of Science, Xi'an Jiaotong University, Xi'an 710049, China
[8] Independent Researcher, Fastlinger Strasse 17, 80999 München, Germany
[9] Physics Department, University of Rijeka, 51000 Rijeka, Croatia
[10] Physics Department, Messiah University, Mechanicsburg, PA 17055, USA
[11] Department of Physics and Astronomy, University of Manchester, Manchester, UK
[12] Physics Department, Boğaziçi University, 34342 Istanbul, Türkiye
[13] Center for Axion and Precision Physics Research, Institute for Basic Science, Daejeon 34051, Korea
[14] Department of Physics, Korea Advanced Institute of Science and Technology, Daejeon 34141, Korea

[#] Present address: University of Hamburg, 20146 Hamburg, Germany.

[&] Corresponding author e-mail: marios.maroudas@cern.ch

## Abstract

The manifestation of the dark Universe began with unexpected large-scale astronomical observations. We are investigating the possible origin of small-scale anomalies, like the annual stratospheric temperature anomalies. Unexpectedly within known physics, their observed planetary relationship does not match concurrent solar activity (F10.7 and EUV emission), whose impact on the atmosphere is unequivocal; this different behavior points to an additional energy source of exo-solar origin. A viable concept behind such observations is based on possible gravitational focusing by the Sun and its planets towards the Earth of low-speed invisible (streaming) matter; its influx towards the Earth gets temporally enhanced. Only a somehow "strongly" interacting invisible streaming matter with the little screened upper atmosphere can be behind the temperature excursions. Ordinary dark matter (DM) candidates like axions or WIMPs, cannot have any noticeable impact. The associated energy deposition $O(\sim W/m^2)$ varies over the 11-year solar cycle. For the widely assumed picture of a quasi-not-interacting DM, the exo-solar energy is enormous. The atmosphere has been uninterruptedly monitored for decades. Therefore, it can serve as a novel (low-threshold) detector for the dark Universe, with built-in spatiotemporal resolution while the Sun's gravity acts temporally as a signal amplifier. Analyzing observations from the anomalous ionosphere we arrived in this work to a surprising relationship with the inner earth activity like earthquakes. Similarly investigating the transient sudden stratospheric warmings within the same reasoning, the nature of the assumed "invisible" streams could be deciphered.

## 1 Introduction

Dark matter (DM) dominates the Universe and came from long-range gravitational observations. Following ongoing searches, DM does not interact with ordinary matter, at least for the parameter phase space such searches are sensitive to. Though, on much smaller scales, a number of unexpected phenomena contradict this global picture for DM (Zioutas et al., 2020). Here we refer to the possible origin of small-scale anomalies, like that of the annually observed temperature excursions in the upper stratosphere (38.5-47.5 km). The observed planetary relationships of the daily stratospheric temperature distribution are unexpected within known physics. We stress that following known physics a remote planetary impact is extremely feeble (Javaraiah, J., 2003) and cannot cause a visible interaction. Interestingly, the observed spectral shapes of the stratospheric temperatures do not match concurrent solar activity (given by the proxy F10.7 radio line at 2.8 GHz), or Sun's EUV emission, whose impact on the atmosphere is unequivocal; this remarkable behaviour points to an additional energy source of exo-solar origin whose energy deposition at the stratosphere is comparable with that incident from our Sun (Zioutas et al., 2020).

Notably, a viable concept behind such observations is based on possible gravitational focusing and/or self-focusing effects by the Sun and its planets towards the Earth of low-speed invisible (streaming) matter, including planetary intrinsic self-focusing effects (Sofue Y., 2003). For example, when the Sun–Earth or Moon-Earth directions align with an invisible stream, its influx towards the Earth's atmosphere can get temporally enhanced by orders of magnitude. This is the key process behind the reasoning of this work.

We denote generic constituents from the dark Universe as "invisible matter", in order to distinguish them from ordinary DM candidates like axions or WIMPs, which cannot have any noticeable impact. Moreover, the observed peaking planetary relations exclude on their own any conventional explanation or coming from the dynamics of the inner atmosphere. Only a somehow "strongly" interacting invisible streaming DM with the little screened upper stratosphere (overhead$\approx$1 gr/cm$^2$ or even much less towards the ionosphere) can be inevitably behind the occasionally observed dynamical behaviour of the upper atmosphere. We recall that candidates from the dark sector are already discussed in the literature. For instance, Anti Quark Nuggets (AQNs) or magnetic monopoles or dark photons are presently the unbiased candidates. Such particles are inspiring this type of work.

The associated energy deposition is O($\sim$W/m$^2$) (Zioutas et al., 2020) is variable over the 11-years solar cycle. For the widely assumed picture of a quasi-not-interacting dark Universe, the new exo-solar energy is enormous. Noticeably, our observationally derived conclusions are not in conflict with the null results of all underground dark matter experiments, given that a similar planetary relationship is not observed even underneath the stratosphere (16–31 km). Interestingly, the atmosphere is uninterruptedly monitored since decades. Therefore, it can serve also parasitically as a novel (low threshold) detector for the dark Universe, with built-in spatiotemporal resolution and the Sun acting temporally as signal amplifier due to its gravitational focusing effects. Observations, for example, like the transient sudden stratospheric warmings, or the anomalous ionosphere, could help to decipher the nature of the putative "invisible streams". With this contribution to COMECAP 2021 we want to communicate this novel idea to atmospheric/climate experts aiming for interesting feedback. The main reasoning of this contribution is based on results from the recent publication (Zioutas et al., 2020). We also include in this manuscript first recent result which correlates the dynamical behaviour of the global Ionosphere with the activity of the inner Earth. More striking observations will be shown in the presentation during the COMECAP 2021 conference. Presently more possibilities are being scrutinized, (see for example https://arxiv.org/abs/2012.03353).

## 2 Previous results

The most relevant results about the stratosphere's dynamical behaviour are given in Zioutas et al., (2020). A few plots are added here to help the reader to follow the reasoning behind the present work. For example:

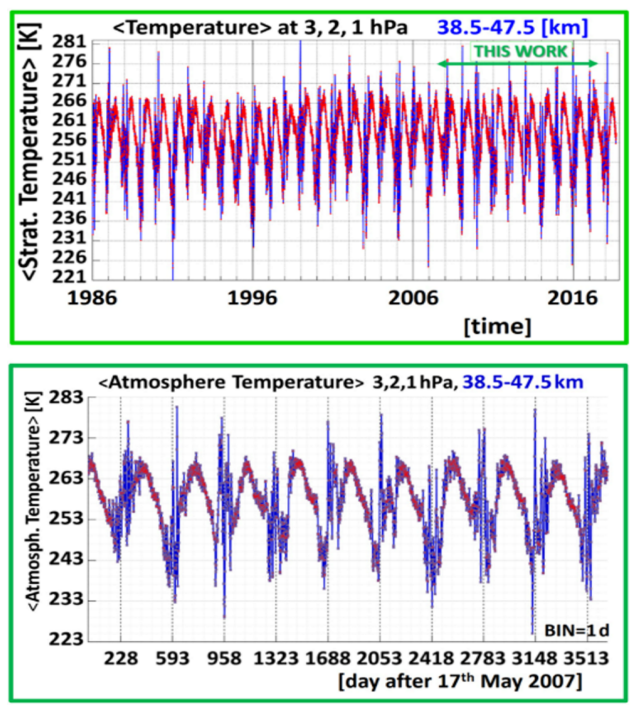

**Fig. 1.** (Top) Time dependence of the mean daily stratospheric temperature [(00:00+12:00)/2] at 3, 2, 1 hPa (altitude ≈ 38.5, 42.5, 47.5 km), 42.5ºN/13.5ºE and for the period 1986–2018. In the bottom is shown the period 2007-2017. The error bar of each individual observation is equal to 0.5 K (Zioutas et al., 2020).

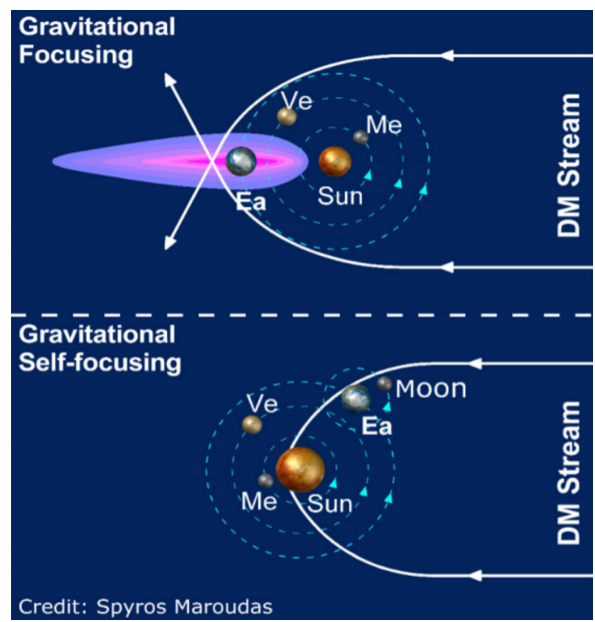

**Fig. 2.** Cartoon illustration of gravitational (self-)focusing effects of DM streams by the Sun, Earth, Venus, Mercury and/or Moon. Top: Gravitational focusing effect by the solar system. In this configuration, the galactic center is on the right side and in the opposite direction of the incident DM stream. Bottom: The self-focusing effect of incident low speed streams reflects the dominating free fall towards the Sun, whose flux towards the Earth can also be gravitationally modulated by the Moon (Javaraiah, J., 2003). The additional gravitational focusing effects due to the inner Earth mass distribution (Javaraiah, J., 2003) can be enormous (not shown).

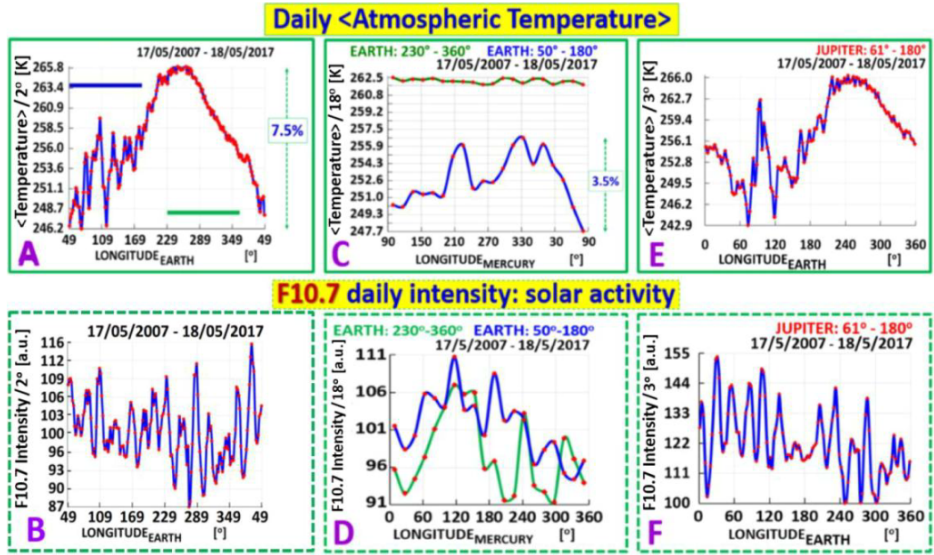

**Fig. 3.** (Top) Planetary longitudinal distributions for the mean daily temperature [(00:00+12:00)/2] of the upper stratospheric layers at 3, 2, 1 hPa (altitude ≈ 38.5, 42.5, 47.5 km), 42.5ºN/13.5ºE and for the period 2007–2017. The blue and green bars in (A) give the Earth's orbital constraints used in (C), which shows planetary relationship for Mercury's reference frame only if the Earth propagates in the heliocentric orbital arcs 50º to 180º. Also, the plot in (E) shows a clear planetary relationship. (Bottom, inside dashed frames) The spectra (B), (D) and (F) show the corresponding longitudinal distributions of the F10.7 solar line (≈2.8 GHz), which is a proxy for the solar activity. A comparison between (A) and (B), (C) and (D), (E) and (F) shows no similarity between each pair. This excludes that the solar radiation is solely at the origin of the upper panels (see also Figure 4).

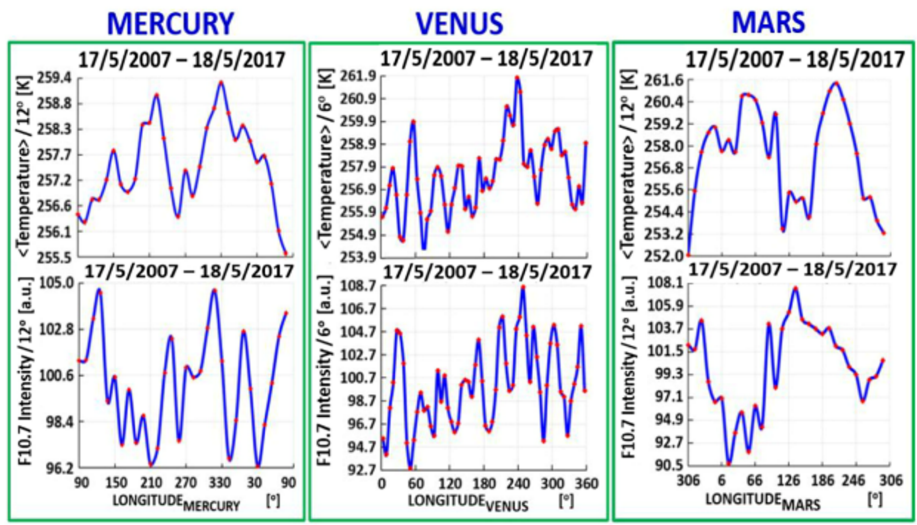

**Fig. 4.** The upper spectra give the mean daily temperature distribution measured at 3,2,1 hPa (altitude ≈38.5–47.5 km) and at the location 42.5ºN/13.5ºE as a function of Mercury, Venus and Mars longitude. For comparison, the lower spectra show the concurrent solar activity derived from the daily intensity of the F10.7 solar line for the same time interval (2007–2017). The error bars can barely be seen. For Mercury and Mars the spectral shapes are clearly different, while for Venus this is less pronounced. Along with the highly different Earth's spectral shapes, the comparison temperature vs. F10.7 line in this Figure further excludes known solar radiation being the only driving source behind the stratospheric temperature distributions.

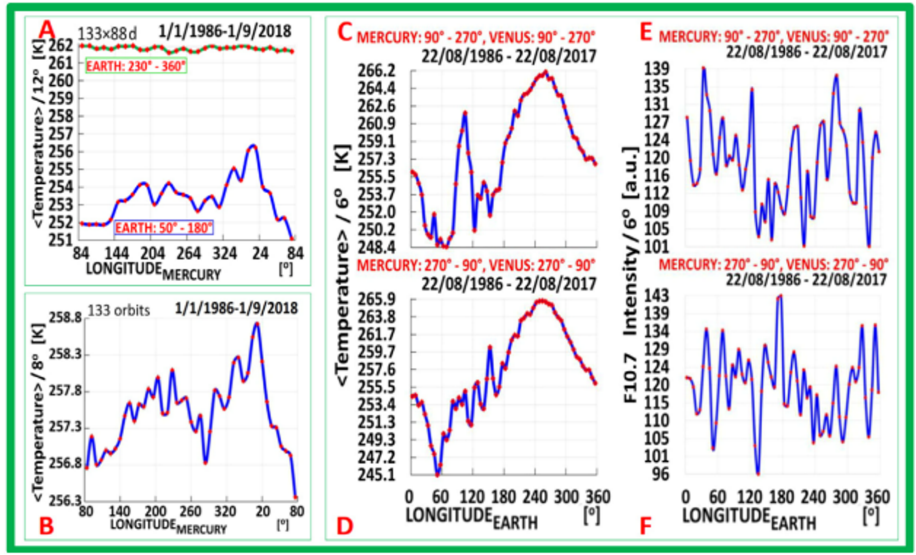

**Fig. 5.** The various spectra are from the upper stratosphere (3,2,1 hPa) and the measuring period is 1986–2018. Spectra E & F, obtained with the F10.7 cm line, are strikingly different from the upper stratospheric temperature distributions C & D, also discarding an involvement of the solar activity exclusively. Concerning the orbital constraints for Mercury and Venus, spectra (C) and (D) are complementary to each other (180o opposite), confirming the observed strongest planetary relationship of the peaking upper stratospheric temperature covering 31 years. Mercury and Earth performed 133 and 31 orbits, respectively.

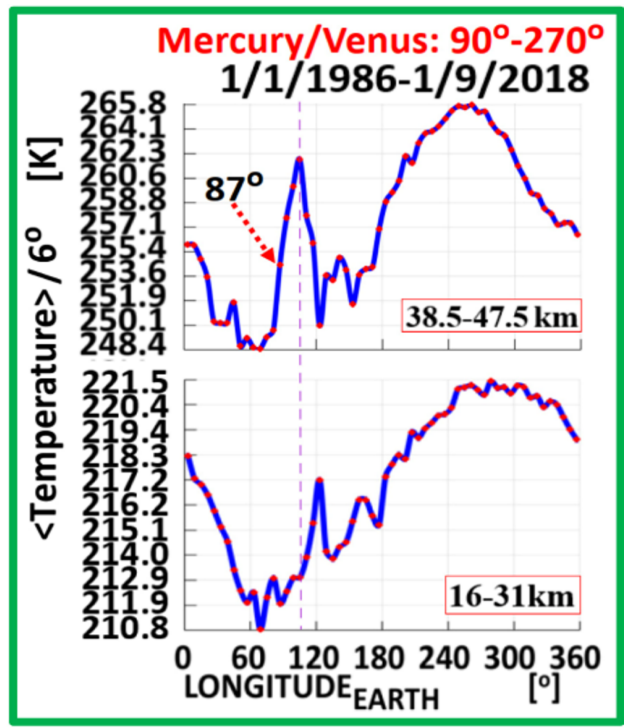

**Fig. 6.** A comparison between the mean temperature spectra of the upper stratosphere (top) and the lower stratosphere (bottom). The lower stratosphere (16–31 km) is the main Ozone layer, which is strongly affected by the solar UV. Of note, the striking difference between both spectra. The position of the Galactic Center in this plot is at ~86.5o, and the upper stratosphere reaches its maximum temperature ~18 days later. The Earth longitude of 100o corresponds to 1st of January.

## 2.1 Correlation between dynamical atmosphere and inner earth activity?

The streaming DM scenario was the driving idea of previous work (Zioutas et al., 2020). In this work we have searched for possible correlations between the global activity of the ionosphere and the actually unpredictable activity of the inner Earth. For this, we have compared data from the global ionosphere and the inner Earth, concentrating on large Earthquakes. Figure 7 shows the daily derived value of the total electron content of the ionosphere (TECUs) for a large period around the zero time, as it is defined by the time of occurring the strong Earthquakes. The obtained correlation is apparent.

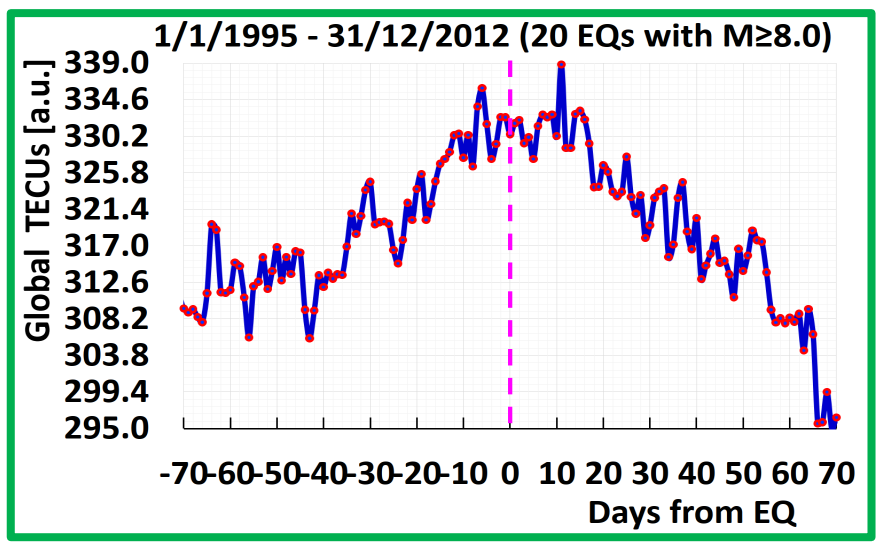

**Fig. 7.** The global degree of ionization of the entire Earth's ionosphere with reference to the observed 20 Earthquakes (EQs) of magnitude M ≥ 8. [Maroudas M., 2021].

## 3 Discussion and Conclusions

The derived correlation between the ionospheric plasma density and the Earthquakes fits-in the picture of streaming DM which interacts on its way towards the earth with the most outer atmospheric layer, i.e., the ionosphere, and the inner earth triggering even a large earthquake. More quantitative 2D plots of the spatiotemporal atmospheric activation the period around a large earthquake will be presented during the conference. Possible candidates from the dark sector are antiquark nuggets (AQNs), magnetic monopoles, and DM dark photons, or others as yet unpredicted theoretically (Zioutas et al., 2020). Their impact can be strongly enhanced by the gravitational lensing effects by the solar system bodies including of course the Sun and even the nearby Moon.

<u>Note added to the conference proceedings</u>: Following Fig. 7 it is obvious that the plasma density of the whole ionosphere can be a warning precursor for very large Earthquakes of about 1 month in advance. This, combined with local GPS measurements could help localize Earthquake occurrences. A similar distribution can be derived for less violent Earthquakes. For this, the same data of this work can be used.

## Acknowledgments

For Y. K. Semertzidis this work was supported by IBS-R017-D1. For M. Maroudas, this research is co-financed by Greece and the European Union (European Social Fund- ESF) through the Operational Programme «Human Resources Development, Education and Lifelong Learning» in the context of the project "Strengthening Human Resources Research Potential via Doctorate Research – 2nd Cycle" (MIS-5000432), implemented by the State Scholarships Foundation (IKY).